\documentclass[12pt,letterpaper]{article}
\usepackage{epsfig,setspace,amsmath,epsf,amssymb,bm,theorem,cite, graphicx, epstopdf, algorithm, algpseudocode,float,color}
\usepackage{multirow}
\usepackage[table,xcdraw]{xcolor}

\newtheorem{theorem}{Theorem}

\newtheorem{remark}{Remark}
\newtheorem{lemma}{Lemma}

\newenvironment{Proof}[1]{\medskip\par\noindent{\bf Proof:\,}\,#1}{{\mbox{\,$\blacksquare$}\par}}

\allowdisplaybreaks

\begin{document}
	
\title{The Capacity of Private Information Retrieval with Partially Known Private Side Information\thanks{This work was supported by NSF Grants CNS 13-14733, CCF 14-22111, CNS 15-26608 and CCF 17-13977.}}
	
\author{Yi-Peng Wei \qquad Karim Banawan \qquad Sennur Ulukus\\
	\normalsize Department of Electrical and Computer Engineering\\
	\normalsize University of Maryland, College Park, MD 20742 \\
	\normalsize {\it ypwei@umd.edu  \qquad \it kbanawan@umd.edu} \qquad {\it ulukus@umd.edu}}
	
\maketitle
	
\vspace*{-0.4cm}

\begin{abstract}
We consider the problem of private information retrieval (PIR) of a single message out of $K$ messages from $N$ replicated and non-colluding databases where a cache-enabled user (retriever) of cache-size $M$ possesses side information in the form of full messages that are partially known to the databases. In this model, the user and the databases engage in a two-phase scheme, namely, the prefetching phase where the user acquires side information and the retrieval phase where the user downloads desired information. In the prefetching phase, the user receives $m_n$ full messages from the $n$th database, under the cache memory size constraint $\sum_{n=1}^N m_n \leq M$. In the retrieval phase, the user wishes to retrieve a message such that no individual database learns anything about the identity of the desired message. In addition, the identities of the side information messages that the user did not prefetch from a database must remain private against that database. Since the side information provided by each database in the prefetching phase is known by the providing database and the side information must be kept private against the remaining databases, we coin this model as \textit{partially known private side information}. We characterize the capacity of the PIR with partially known private side information to be $C=\left(1+\frac{1}{N}+\cdots+\frac{1}{N^{K-M-1}}\right)^{-1}=\frac{1-\frac{1}{N}}{1-(\frac{1}{N})^{K-M}}$. Interestingly, this result is the same if none of the databases knows any of the prefetched side information, i.e., when the side information is obtained externally, a problem posed by Kadhe et al.~and settled by Chen-Wang-Jafar recently. Thus, our result implies that there is no loss in using the same databases for both prefetching and retrieval phases.
\end{abstract}	
	
\section{Introduction}
The private information retrieval (PIR) problem is a canonical problem to study privacy issues that arise when information is downloaded (retrieved) from public databases. Since its first formulation by Chor et al.~in \cite{ChorPIR}, the PIR problem has become a central research topic in the computer science literature, see e.g., \cite{PIRsurvey2004, cachin1999computationally, ostrovsky2007survey, yekhanin2010private}. In the classical setting of PIR in \cite{ChorPIR}, a user wishes to retrieve a single message (or a file) out of $K$ messages replicated across $N$ non-communicating databases without leaking any information about the identity of the retrieved message. To that end, the user submits a query to each database. Each database responds truthfully with an answer string. The user reconstructs the desired message from the collected answer strings. Trivially, the user can download the entire database and incur a linear (in number of messages) download cost, but this retrieval strategy is highly inefficient. The efficiency of a PIR scheme is measured by the normalized download cost, which is the cost of privately downloading one bit of the desired message. The goal of the PIR problem is to devise the most efficient retrieval strategy under the privacy and decodability constraints.

The PIR problem has received attention in recent years in the information and coding theory literatures, see e.g., \cite{RamchandranPIR, unsynchonizedPIR, YamamotoPIR, VardyConf2015, RazanPIR, JafarConf2016}. In the leading work of Sun-Jafar \cite{JafarPIR}, the classical PIR problem is re-formulated to conform with the conventional information-theoretic arguments, and the notion of PIR capacity is introduced, which is defined as the supremum of retrieval rates over all achievable retrieval schemes. Reference \cite{JafarPIR} characterizes the capacity of the classical PIR model to be $C=\left(1+\frac{1}{N}+\cdots+\frac{1}{N^{K-1}}\right)^{-1}$ using a greedy achievable scheme that is closely related to blind interference alignment \cite{JafarPIRBlind} and an induction-based converse argument. Following the work of Sun-Jafar \cite{JafarPIR}, the capacity of many interesting variants of the classical PIR model have been investigated, such as, PIR from colluding databases, robust PIR, symmetric PIR, PIR from MDS-coded databases, PIR for arbitrary message lengths, multi-round PIR, multi-message PIR, PIR from Byzantine databases, secure symmetric PIR with adversaries, and their several combinations  \cite{JafarColluding, arbitraryCollusion, symmetricPIR, KarimCoded, arbmsgPIR, MultiroundPIR, MPIRjournal, BPIRjournal, codedsymmetric, codedcolluded, codedcolludedJafar, codedcolludingZhang, MPIRcodedcolludingZhang, wang2017linear, symmetricByzantine}.

In this paper, we consider the problem of PIR with partially known private side information. Our work is most closely related to \cite{tandon2017capacity, kadhe2017private, wei2017fundamental, chen2017capacity}\footnote{A parallel line of work that studies privacy issues of requests and side information in index coding based broadcast systems can be found in \cite{privateBC_karmoose, k_limited_karmoose}.}. These works investigate the PIR problem when the user (retriever) possesses some form of side information about the contents of the databases. However, the models of \cite{tandon2017capacity, kadhe2017private, wei2017fundamental, chen2017capacity} differ in three important aspects, namely, 1) the structure of the side information, 2) the presence or absence of privacy constraints on the side information, and 3) the databases' awareness of the side information at its initial acquisition. Here, structure of the side information refers to whether the side information is in the form of full messages or parts of messages or whether messages are mixed through functions (coded/uncoded); privacy of the side information refers to whether the user further aims to keep the side information private from the databases; and databases' awareness of the side information refers to whether the databases knew the initially prefetched side information.

Specifically, reference \cite{tandon2017capacity} studies the capacity of the cache-aided PIR where the user caches $rLK$ bits in the form of any arbitrary function of the $K$ messages, where $L$ is the message size, and $0 \leq r\leq 1$ is the caching ratio. Reference \cite{tandon2017capacity} assumes that the cache content is perfectly known by all the databases, and hence there is no need to protect the privacy of the cached content. Reference \cite{tandon2017capacity} determines the optimal download cost for this model to be $D^*(r)=\frac{1}{C(r)}= (1-r)\left(1+\frac{1}{N}+\cdots+\frac{1}{N^{K-1}}\right)$ using a memory-sharing achievable scheme and a converse that utilizes Han's inequality. This conclusion is somewhat pessimistic in that the user cannot exploit the cached content as useful side information during PIR to reduce the download cost, since the databases are fully aware of it; the optimum $D^*(r)$ formula indicates that the user should download the uncached part of the content, i.e., $(1-r)$, via the optimum PIR scheme in \cite{JafarPIR}. This result motivates \cite{kadhe2017private, wei2017fundamental} to study the other extreme when the databases are completely unaware of the side information at its initial acquisition. References \cite{kadhe2017private} and \cite{wei2017fundamental} differ in terms of the structure of the cached content: \cite{kadhe2017private} considers the case where $r K$ full messages are cached, and \cite{wei2017fundamental} considers the case where a random $r$ fraction of the symbols of each of $K$ messages is cached. In this case, \cite{wei2017fundamental} shows a significant reduction in the download cost over \cite{tandon2017capacity}, as the user can now leverage the cached bits as side information, since they are unknown to the databases. In \cite{wei2017fundamental}, there is no privacy constraint on the cached content.

Reference \cite{kadhe2017private} further introduces another model where the cached content (in the form of full messages) which is unknown to the databases at the time of initial prefetching, must remain unknown throughout the PIR, i.e., the queries of the user should not leak any information about the cached content to the databases. The exact capacity for this problem is settled in \cite{chen2017capacity} to be $C=\left(1+\frac{1}{N}+\cdots+\frac{1}{N^{K-M-1}}\right)^{-1}$. The optimal achievable scheme in this case starts from the traditional achievable scheme without side information in \cite{JafarPIR} and reduces the download cost by utilizing the reconstruction property of MDS codes.

In this paper, we take a deeper look at the issue of {\it awareness} or otherwise {\it unawareness} of the databases about the cached content {\it at its initial acquisition}. We first note that it is practically challenging to make the side information completely unknown to the databases at its initial acquisition as assumed in \cite{wei2017fundamental, kadhe2017private, chen2017capacity}. One way to do this could be to employ one of the databases for prefetching the side information and exclude it from the retrieval process. Therefore, for the remaining $N-1$ databases, the side information is completely unknown. This solution is strictly sub-optimal as the capacity expression in \cite{chen2017capacity} (shown as $C$ in the previous paragraph) is monotonically decreasing in $N$. An alternative solution could be to devise a refreshing mechanism that ensures that the cached content is essentially random from the perspective of each database \cite{tandon2017capacity}, which may be challenging to implement. We also note that the other extreme of the problem, where the databases are fully aware of the cached content \cite{tandon2017capacity}, is discouraging as the user cannot benefit from the cached side information. Therefore, a natural model is to use the databases for both prefetching and retrieval phases, such that the databases gain partial knowledge about the side information available to the user, which makes it possible for the user to exploit the remaining side information that is unknown to each individual database to reduce the download cost during the retrieval process. This poses the following questions: Can we propose efficient joint prefetching-retrieval strategies that exploit the limited knowledge of each database to drive down the download cost? How much is the loss from the fully unknown case in \cite{kadhe2017private, chen2017capacity}?

In this paper, we investigate the PIR problem when the user and the databases engage in a two-phase scheme, namely, prefetching phase and retrieval phase. In the prefetching phase, the user caches $m_n$ full messages out of the $K$ messages from the $n$th database under a total cache memory size constraint $\sum_{n=1}^N m_n \leq M$. Hence, each database has a \textit{partial knowledge} about the side information possessed by the user, namely, the part of the side information that this database has provided during the prefetching phase. In the retrieval phase, the user wants to retrieve a message (which is not present in its memory) without leaking any information to any individual database about the desired message or the remaining side information messages that are unknown to each database. The goal of this work is to design a joint prefetching-retrieval scheme that minimizes the download cost in the retrieval phase.

To that end, we first derive a general lower bound for the normalized download cost that is independent of the prefetching strategy. Then, we prove that this bound is attainable using two achievable schemes. The first achievable scheme, which is proposed in \cite{chen2017capacity} for completely unknown side information, is a valid achievable scheme for our problem with partially known side information for any prefetching strategy.\footnote{We thank Dr. Hua Sun for pointing this out.} We provide a second achievable scheme for the case of uniform prefetching, i.e., $m_n=\frac{M}{N} \in \mathbb{N}$, which requires smaller sub-packetization and smaller field size for realizing MDS codes. While the first achievable scheme \cite{chen2017capacity} requires a message size of $L=N^{K}$, the second achievable scheme proposed here requires a message size of $L=N^{K-\frac{M}{N}}$, which scales down the message size by an exponential factor $N^\frac{M}{N}$, which in turn simplifies the achievable scheme and minimizes the total number of downloaded bits without sacrificing from the capacity. We prove that the exact capacity of this problem is $C=\left(1+\frac{1}{N}+\cdots+\frac{1}{N^{K-M-1}}\right)^{-1}$. Surprisingly, this is the same capacity expression for the PIR problem when the databases are completely unaware of the side information possessed by the user as found in \cite{chen2017capacity} recently. Therefore, our result implies that there is no loss in the capacity if the same databases are employed in both prefetching and retrieval phases.

\section{System Model}
We consider a classic PIR problem with $K$ independent messages $W_1, \dots, W_K$, where each message consists of $L$ symbols,
\begin{align}
H(W_1)=\dots=H(W_K)=L, \qquad H(W_1, \dots, W_K)=H(W_1)+\dots+H(W_K).
\end{align}
There are $N$ non-communicating databases, and each database stores all the $K$ messages. The user (retriever) has a local cache memory which can store up to $M$ messages.

There are two phases: a \textit{prefetching phase} and a \textit{retrieval phase}. In the prefetching phase, $\forall n\in [N]$, where $[N]= \{1,2, \dots, N\}$, the user caches $m_n$ out of total $K$ messages from the $n$th database. We denote the indices of the cached messages from the $n$th database as $\mathbb{H}_{n}$. Therefore, $|\mathbb{H}_{n}|=m_n$. We denote the indices of all cached messages as $\mathbb{H}$,
\begin{align}
\mathbb{H}= \bigcup_{n=1}^N \mathbb{H}_n,
\end{align}
where $\mathbb{H}_{n_1} \cap \mathbb{H}_{n_2} = \emptyset$, if $n_1 \neq n_2$. Due to the cache memory size constraint, we require
\begin{align} \label{memory_size_constraint}
|\mathbb{H}|=\sum_{n=1}^{N} m_n \leq M.
\end{align}
Since the user caches $m_n$ messages from the $n$th database, $\mathbb{H}_n$ is known to the $n$th database. Since the databases do not communicate with each other, $\mathbb{H}_n$ is unknown to the other databases. We use $\mathbf{m}=(m_1,\dots, m_N)$ to represent the prefetching phase. After the prefetching phase, the user learns $|\mathbb{H}|$ messages, denoted as $\mathcal{W}_{\mathbb{H}}=\{ W_{i_1}, \dots, W_{i_{|\mathbb{H}|}} \}$. We refer to $\mathcal{W}_{\mathbb{H}}$ as \textit{partially known private side information}.

In the retrieval phase, the user privately generates a desired message index $\theta \in [K] \setminus \mathbb{H}$, and wishes to retrieve message $W_\theta$ such that no database knows which message is retrieved. Since the desired message index $\theta$ and cached message indices $\mathbb{H}$ are independent of the message contents, for random variables $\theta$, $\mathbb{H}$, and $W_1,\dots,W_K$, we have
\begin{align} \label{independency}
H\left(\theta, \mathbb{H}, W_1,\dots,W_K  \right)= H\left( \theta, \mathbb{H} \right) + H(W_1)+\dots+H(W_K).
\end{align}

In order to retrieve $W_\theta$, the user sends $N$ queries $Q_1^{[\theta, \mathbb{H}]}, \dots, Q_N^{[\theta, \mathbb{H}]}$ to the $N$ databases, where $Q_n^{[\theta, \mathbb{H}]}$ is the query sent to the $n$th database for message $W_\theta$ given the user has  partially known private side information $\mathcal{W}_{\mathbb{H}}$. The queries are generated according to $\mathbb{H}$, which is independent of the realizations of the $K$ messages. Therefore, we have
\begin{align} \label{query_indep}
I(W_1, \dots, W_K; Q_1^{[\theta, \mathbb{H}]}, \dots,  Q_N^{[\theta, \mathbb{H}]}  ) =0.
\end{align}

To ensure that individual databases do not know which message is retrieved and also do not know the cached messages from other databases, i.e., to guarantee the privacy of $(\theta, \mathbb{H} \setminus \mathbb{H}_n)$, we need to satisfy the following privacy constraint, $\forall n \in [N]$, $\forall \mathbb{H}, \mathbb{H}'$ such that $|\mathbb{H}|=|\mathbb{H}'| \leq M$, $\mathbb{H}_n \subset \mathbb{H}$, $\mathbb{H}_n \subset \mathbb{H}'$, and $\forall \theta \in [K]\setminus \mathbb{H}$, $\forall \theta' \in [K]\setminus \mathbb{H}'$,
\begin{align} \label{privacy_constraint}
(Q_n^{[\theta,\mathbb{H}]}, A_n^{[\theta,\mathbb{H}]}, W_1, \dots, W_K, \mathbb{H}_n)
\sim (Q_n^{[\theta',\mathbb{H}']}, A_n^{[\theta',\mathbb{H}']}, W_1, \dots, W_K, \mathbb{H}_n),
\end{align}
where $A \sim B$ means that $A$ and $B$ are identically distributed.

Upon receiving the query $Q_n^{[\theta,\mathbb{H}]}$, the $n$th database replies with an answering string $A_n^{[\theta,\mathbb{H}]}$, which is a function of  $Q_n^{[\theta,\mathbb{H}]}$ and all the $K$ messages. Therefore, $\forall \theta \in [K]\setminus \mathbb{H}, \forall n \in [N]$,
\begin{align} \label{answer_constraint}
H(A_n^{[\theta, \mathbb{H}]}|Q_n^{[\theta, \mathbb{H}]}, W_1, \dots, W_K)=0.
\end{align}

After receiving the answering strings $A_1^{[\theta, \mathbb{H}]}, \dots, A_N^{[\theta, \mathbb{H}]}$ from all the $N$ databases, the user needs to decode the desired message $W_\theta$ reliably. By using Fano's inequality, we have the following reliability constraint
\begin{align} \label{reliability_constraint}
H\left(W_\theta|\mathcal{W}_{\mathbb{H}}, \mathbb{H}, Q_1^{[\theta, \mathbb{H}]}, \dots, Q_N^{[\theta, \mathbb{H}]}, A_1^{[\theta, \mathbb{H}]}, \dots, A_N^{[\theta, \mathbb{H}]} \right) = o(L),
\end{align}
where $o(L)$ denotes a function such that $\frac{o(L)}{L} \rightarrow 0$ as $L \rightarrow \infty$.

For fixed $N$, $K$, and pretching scheme $\mathbf{m}=(m_1, \dots, m_N)$, a pair $(D(\mathbf{m}),L(\mathbf{m}))$ is achievable if there exists a PIR scheme for messages of size $L(\mathbf{m})$ symbols long with partially known private side information satisfying the privacy constraint \eqref{privacy_constraint} and the reliability constraint \eqref{reliability_constraint}, where $D(\mathbf{m})$ represents the expected number of downloaded symbols (over all the queries) from the $N$ databases via the answering strings $A_{1:N}^{[\theta, \mathbb{H}]}$, where $A_{1:N}^{[\theta, \mathbb{H}]}=(A_1^{[\theta, \mathbb{H}]}, \dots, A_N^{[\theta, \mathbb{H}]})$, i.e.,
\begin{align}
D(\mathbf{m})=\sum_{n=1}^N H\left(A_n^{[\theta, \mathbb{H}]}\right).
\end{align}
In this work, for fixed $N$, $K$, and $M$, we aim to characterize the optimal normalized download cost $D^*$, where
\begin{align} \label{D*}
D^*= \inf_{\mathbf{m}:\eqref{memory_size_constraint}}  \left\{ \frac{D(\mathbf{m})}{L(\mathbf{m})}: \left(D(\mathbf{m}), L(\mathbf{m}) \right) \text{ is achievable}     \right\}.
\end{align}

\section{Main Results}

We characterize the exact normalized download cost for the PIR problem with partially known private side information as shown in the following theorem.
\begin{theorem}\label{thm1}
In the PIR problem with partially known private side information under the cache memory size constraint $|\mathbb{H}|\leq M$, the optimal normalized download cost is
\begin{align}
D^* &= 1+\frac{1}{N}+\cdots+\frac{1}{N^{K-M-1}} \label{thm1_lb}\\
&=\frac{1-(\frac{1}{N})^{K-M}}{1-\frac{1}{N}}.
\end{align}
\end{theorem}

The converse proof for Theorem~\ref{thm1} is given in Section~\ref{converse}, and the achievability proof for Theorem~\ref{thm1} is given in Section~\ref{achievability}. Theorem~\ref{thm1} does not assume any particular property for the prefetching strategy, i.e., $\mathbf{m}$ is arbitrary except for satisfying the memory size constraint. We have a few remarks.

\begin{remark}
Theorem~\ref{thm1} implies that $C=\frac{1}{D^*}=\frac{1-\frac{1}{N}}{1-(\frac{1}{N})^{K-M}}$. Surprisingly, this capacity expression is exactly the same as the capacity for the PIR problem with completely unknown private side information in \cite{chen2017capacity}. This implies that there is no loss in capacity due to employing the same databases for both prefetching and retrieval phases. The reason for this phenomenon is that although each database has a partial knowledge about some of the cached messages at the user, the privacy constraint on this known side information is relaxed.
\end{remark}

\begin{remark}
The normalized download cost in Theorem~\ref{thm1} is the same as the normalized download cost for the classical PIR problem \cite{JafarPIR} if the number of messages is $K-M$. That is, a cache of size $M$ messages effectively reduces the total number of messages by $M$. Noting that the download cost in \cite{JafarPIR} monotonically increases in the number of messages, the effective reduction in the number of messages by the cache size results in a significant reduction in the download cost due to the presence of side information at the user even though it is partially known by the databases and it needs to be kept private against other databases.
\end{remark}

\begin{remark}
The optimal prefetching strategy exploits the entire cache memory of the user as the capacity expression is monotonically increasing in $M$.
\end{remark}

\begin{remark}
In Section~\ref{achievability}, we present the capacity achieving schemes for the partially known private side information. We note that, in general the PIR scheme in \cite{chen2017capacity} is a valid achievable scheme for our problem as well. Nevertheless, in the special case of \emph{uniform prefetching}, i.e., $m_n=\frac{M}{N}=m \in \mathbb{N}$, we provide a different achievable scheme that exploits the prefetching uniformity to work with message size $L=N^{K-m}=N^{K-\frac{M}{N}}$ in contrast to $L=N^{K}$ needed for the scheme in \cite{chen2017capacity}, i.e., the message size is decreased by an exponential factor $N^\frac{M}{N}$. Furthermore, we note that although both schemes need an MDS code to reduce the number of downloaded equations, we note that the field size needed to realize this MDS code is significantly smaller with our scheme (if $\frac{M}{N} \in \mathbb{N}$) compared with the field size needed in the scheme in \cite{chen2017capacity}. This implies that although \emph{uniform prefetching} does not affect the PIR capacity, it significantly simplifies the achievable scheme.
\end{remark}

\section{Converse Proof} \label{converse}

In this section, we derive a general lower bound for the normalized download cost $D^*$ given in \eqref{D*}. We extend the techniques presented in \cite{JafarPIR,chen2017capacity} to the PIR problem with partially known private side information.

For the prefetching vector $\mathbf{m}=(m_1,\dots, m_N)$ satisfying \eqref{memory_size_constraint}, we note that satisfying the memory size constraint with equality leads to a valid lower bound on \eqref{D*}. Consequently, we first consider the case $\sum_{n=1}^N m_n=\tilde{M}\leq M$, i.e., we study the case when the user learns $\tilde{M}$ messages after the prefetching phase. Since we do not specify the prefetching strategy $\mathbf{m}$ in advance, the following lower bound is valid for all $\mathbf{m}$ such that $\sum_{n=1}^N m_n=\tilde{M}$. Without loss of generality, we relabel the $\tilde{M}$ cached messages as $W_1, W_2, \dots, W_{\tilde{M}}$, i.e., $\mathbb{H}=\{1,2, \dots, \tilde{M}\}$ and $\mathcal{W}_{\mathbb{H}}=W_{1:\tilde{M}}$. We first need the following lemma, which characterizes a lower bound on the length of the undesired portion of the answering strings as a consequence of the privacy constraint on the retrieved message.

\begin{lemma}[Interference lower bound]\label{lemma_converse1}
For the PIR with partially known private side information, the interference from undesired messages within the answering strings, $D-L$, is lower bounded by, 	
\begin{align} \label{eq_L1}
D -L + o(L) \geq I\left(W_{ \tilde{M}+2:K}; \mathbb{H}, Q_{1:N}^{[ \tilde{M}+1, \mathbb{H}]}, A_{1:N}^{[ \tilde{M}+1, \mathbb{H}]}|\mathcal{W}_{\mathbb{H}}, W_{\tilde{M}+1} \right).
\end{align}
\end{lemma}

If the privacy constraint is absent, the user downloads only $L$ symbols for the desired message, however, when the privacy constraint is present, it should download $D$ symbols. The difference between $D$ and $L$, i.e., $D-L$, corresponds to the undesired portion of the answering strings. Note that Lemma~\ref{lemma_converse1} is an extension of \cite[Lemma~5]{JafarPIR} if $\tilde{M}=0$, i.e., the user has no partially known private side information. Lemma~\ref{lemma_converse1} differs from its counterpart in \cite[Lemma~1]{wei2017fundamental} in two aspects, namely, the left hand side is $D(r)-L(1-r)$ in \cite{wei2017fundamental} as the user requests to download the uncached bits only, and the bound in \cite[Lemma~1]{wei2017fundamental} constructs $K-1$ distinct lower bounds by changing $k$ in contrast to one bound here as it always starts from $W_{\tilde{M}+2}$. Finally, we note that a similar argument to Lemma~\ref{lemma_converse1} can be implied from \cite{chen2017capacity}.

\begin{Proof}
We start with the right hand side of \eqref{eq_L1},	
\begin{align}
&I\left(W_{\tilde{M}+2:K};\mathbb{H},Q_{1:N}^{[\tilde{M}+1, \mathbb{H}]}, A_{1:N}^{[\tilde{M}+1, \mathbb{H}]} |\mathcal{W}_{\mathbb{H}}, W_{\tilde{M}+1} \right) \notag \\
&\qquad=I\left(W_{\tilde{M}+2:K}; \mathbb{H}, Q_{1:N}^{[\tilde{M}+1, \mathbb{H}]}, A_{1:N}^{[\tilde{M}+1, \mathbb{H}]}, W_{\tilde{M}+1} |\mathcal{W}_{\mathbb{H}} \right)
- I \left(W_{\tilde{M}+2:K}; W_{\tilde{M}+1}|\mathcal{W}_{\mathbb{H}}  \right). \label{eq_lemma_1}
\end{align}	
For the first term on the right hand side of \eqref{eq_lemma_1}, we have
\begin{align}
& I\left(W_{\tilde{M}+2:K}; \mathbb{H}, Q_{1:N}^{[\tilde{M}+1, \mathbb{H}]}, A_{1:N}^{[\tilde{M}+1,\mathbb{H}]}, W_{\tilde{M}+1}|\mathcal{W}_{\mathbb{H}}\right) \notag \\
&~=I\left(W_{\tilde{M}+2:K};\mathbb{H},Q_{1:N}^{[\tilde{M}+1,\mathbb{H}]}, A_{1:N}^{[\tilde{M}+1, \mathbb{H}]}|\mathcal{W}_{\mathbb{H}} \right)+ I \left( W_{\tilde{M}+2:K}; W_{\tilde{M}+1}|\mathbb{H}, Q_{1:N}^{[\tilde{M}+1, \mathbb{H}]}, A_{1:N}^{[\tilde{M}+1, \mathbb{H}]},\mathcal{W}_\mathbb{H} \right)  \\
&\label{eq_ILB_1}~\stackrel{\eqref{reliability_constraint}}{=}
I\left(W_{\tilde{M}+2:K};\mathbb{H},Q_{1:N}^{[\tilde{M}+1,\mathbb{H}]}, A_{1:N}^{[\tilde{M}+1, \mathbb{H}]} |\mathcal{W}_\mathbb{H} \right)+ o(L) \\
&\label{eq_ILB_2} \stackrel{\eqref{independency},\eqref{query_indep}}{=}
I\left(W_{\tilde{M}+2:K}; A_{1:N}^{[\tilde{M}+1, \mathbb{H}]} |\mathcal{W}_{\mathbb{H}}, \mathbb{H}, Q_{1:N}^{[\tilde{M}+1,\mathbb{H}]} \right) + o(L) \\
&~= H\left( A_{1:N}^{[\tilde{M}+1, \mathbb{H}]} |\mathcal{W}_{\mathbb{H}}, \mathbb{H}, Q_{1:N}^{[\tilde{M}+1,\mathbb{H}]} \right)
- H\left(A_{1:N}^{[\tilde{M}+1, \mathbb{H}]} |\mathcal{W}_\mathbb{H}, \mathbb{H}, Q_{1:N}^{[\tilde{M}+1,\mathbb{H}]},W_{\tilde{M}+2:K}  \right)   + o(L) \\
&\label{eq_ILB_3}~\stackrel{\eqref{reliability_constraint}}{=}
H\left( A_{1:N}^{[\tilde{M}+1, \mathbb{H}]} |\mathcal{W}_\mathbb{H}, \mathbb{H}, Q_{1:N}^{[\tilde{M}+1,\mathbb{H}]} \right)
-H\left(A_{1:N}^{[\tilde{M}+1, \mathbb{H}]}, W_{\tilde{M}+1} |\mathcal{W}_\mathbb{H}, \mathbb{H}, Q_{1:N}^{[\tilde{M}+1,\mathbb{H}]},W_{\tilde{M}+2:K}  \right)  + o(L) \\
&\label{eq_ILB_4}~\leq H\left( A_{1:N}^{[\tilde{M}+1, \mathbb{H}]} |\mathcal{W}_\mathbb{H}, \mathbb{H}, Q_{1:N}^{[\tilde{M}+1,\mathbb{H}]} \right)
-H\left( W_{\tilde{M}+1} |\mathcal{W}_\mathbb{H}, \mathbb{H}, Q_{1:N}^{[\tilde{M}+1,\mathbb{H}]},W_{\tilde{M}+2:K}  \right)  + o(L) \\
&\label{eq_ILB_5}\stackrel{\eqref{independency},\eqref{query_indep}}{=}
H\left( A_{1:N}^{[\tilde{M}+1, \mathbb{H}]} |\mathcal{W}_\mathbb{H}, \mathbb{H}, Q_{1:N}^{[\tilde{M}+1,\mathbb{H}]} \right)
-H\left( W_{\tilde{M}+1} |\mathcal{W}_\mathbb{H}, W_{\tilde{M}+2:K}  \right) + o(L) \\
&\label{eq_ILB_6}~= H\left( A_{1:N}^{[\tilde{M}+1, \mathbb{H}]} |\mathcal{W}_\mathbb{H}, \mathbb{H}, Q_{1:N}^{[\tilde{M}+1,\mathbb{H}]} \right)-L+o(L) \\
&~ \leq H\left( A_{1:N}^{[\tilde{M}+1, \mathbb{H}]} \right) -L +o(L)  \\
&~ \leq D - L+ o(L), \label{eq_l1_1}
\end{align}
where \eqref{eq_ILB_1}, \eqref{eq_ILB_3} follow from the decodability of $W_{\tilde{M}+1}$ given $\left(\mathbb{H}, Q_{1:N}^{[\tilde{M}+1, \mathbb{H}]}, A_{1:N}^{[\tilde{M}+1, \mathbb{H}]},\mathcal{W}_\mathbb{H}\right)$, \eqref{eq_ILB_2} follows from the independence of $W_{\tilde{M}+2:K}$ and $\left(\mathbb{H}, Q_{1:N}^{[\tilde{M}+1, \mathbb{H}]}\right)$, \eqref{eq_ILB_5} follows from the independence of $W_{\tilde{M}+1}$ and $\left(\mathbb{H}, Q_{1:N}^{[\tilde{M}+1, \mathbb{H}]}\right)$, and \eqref{eq_l1_1} follows from the independence bound.

For the second term on the right hand side of \eqref{eq_lemma_1}, we have
\begin{align}
I \left(W_{\tilde{M}+2:K}; W_{\tilde{M}+1}|\mathcal{W}_\mathbb{H}  \right)
&=H\left(W_{\tilde{M}+1}|\mathcal{W}_\mathbb{H}  \right) -H\left(W_{\tilde{M}+1}|\mathcal{W}_\mathbb{H}, W_{\tilde{M}+2:K}\right) \\
&=L-L=0.   \label{eq_l1_2}
\end{align}

Combining \eqref{eq_lemma_1}, \eqref{eq_l1_1}, and \eqref{eq_l1_2} yields \eqref{eq_L1}.
\end{Proof}

In the following lemma, we prove an inductive relation for the mutual information term on the right hand side of \eqref{eq_L1}.
\begin{lemma}[Induction lemma]\label{lemma_converse2}
For all $k\in \{\tilde{M}+2,\dots,K\}$, the mutual information term in Lemma~\ref{lemma_converse1} can be inductively lower bounded as,
\begin{align} \label{eq_L2}
&I\left( W_{k:K} ; \mathbb{H}, Q_{1:N}^{[k-1, \mathbb{H}]}, A_{1:N}^{[k-1, \mathbb{H}]} |\mathcal{W}_\mathbb{H}, W_{\tilde{M}+1:k-1} \right) \notag \\
&\qquad\qquad \geq \frac{1}{N}  I\left(W_{k+1:K};\mathbb{H}, Q_{1:N}^{[k, \mathbb{H}]}, A_{1:N}^{[k, \mathbb{H}]}|\mathcal{W}_\mathbb{H}, W_{\tilde{M}+1:k}  \right)  +\frac{L - o(L)}{N}.
\end{align}	
\end{lemma}

Lemma~\ref{lemma_converse2} is a generalization of \cite[Lemma~6]{JafarPIR} to our setting. The main difference between Lemma~\ref{lemma_converse2} and \cite{chen2017capacity} is that in order to apply the {\it partial} privacy constraint, the random variable $\mathbb{H}$ should be used in its local form $\mathbb{H}_n$ as it corresponds to the partial knowledge of the $n$th database.

\begin{Proof}
We start with the left hand side of \eqref{eq_L2},		
\begin{align}
&I\left( W_{k:K} ; \mathbb{H}, Q_{1:N}^{[k-1, \mathbb{H}]},A_{1:N}^{[k-1,\mathbb{H}]}|\mathcal{W}_\mathbb{H},W_{\tilde{M}+1:k-1}\right) \notag \\
&\qquad=\frac{1}{N} \times N \times I\left(W_{k:K}; \mathbb{H}, Q_{1:N}^{[k-1, \mathbb{H}]}, A_{1:N}^{[k-1, \mathbb{H}]}|\mathcal{W}_\mathbb{H},W_{\tilde{M}+1:k-1}\right) \\
&\qquad \geq \frac{1}{N} \sum_{n=1}^N I\left(W_{k:K}; \mathbb{H}_n, Q_n^{[k-1, \mathbb{H}]}, A_n^{[k-1, \mathbb{H}]}|\mathcal{W}_\mathbb{H},W_{\tilde{M}+1:k-1}\right) \label{eq_L2_1} \\
&\qquad \geq \frac{1}{N} \sum_{n=1}^N I\left(W_{k:K};Q_n^{[k-1, \mathbb{H}]}, A_n^{[k-1, \mathbb{H}]}| \mathcal{W}_\mathbb{H},W_{\tilde{M}+1:k-1} ,\mathbb{H}_n\right) \\
&\qquad \stackrel{\eqref{privacy_constraint}}{=}
\frac{1}{N} \sum_{n=1}^N I\left(W_{k:K};Q_n^{[k, \mathbb{H}]}, A_n^{[k, \mathbb{H}]}| \mathcal{W}_\mathbb{H},W_{\tilde{M}+1:k-1} ,\mathbb{H}_n \right) \label{eq_L2_2}\\
&\quad~\stackrel{\eqref{independency},\eqref{query_indep}}{=}
\frac{1}{N} \sum_{n=1}^N I\left(W_{k:K};A_n^{[k, \mathbb{H}]}|\mathcal{W}_\mathbb{H},W_{\tilde{M}+1:k-1} ,\mathbb{H}_n, Q_n^{[k, \mathbb{H}]} \right) \label{eq_L2_3}\\
&\qquad\stackrel{\eqref{answer_constraint}}{=}
\frac{1}{N}\sum_{n=1}^N H\left( A_n^{[k, \mathbb{H}]}|\mathcal{W}_\mathbb{H},W_{\tilde{M}+1:k-1} ,\mathbb{H}_n, Q_n^{[k, \mathbb{H}]} \right)\label{eq_L2_4}\\
&\qquad \geq \frac{1}{N}\sum_{n=1}^N  H\left( A_n^{[k, \mathbb{H}]}|\mathcal{W}_\mathbb{H},W_{\tilde{M}+1:k-1},\mathbb{H},Q_{1:N}^{[k, \mathbb{H}]}, A_{1:n-1}^{[k, \mathbb{H}]}\right) \label{eq_L2_5}\\
&\qquad \stackrel{\eqref{answer_constraint}}{=}
\frac{1}{N}\sum_{n=1}^N I \left(W_{k:K}; A_n^{[k, \mathbb{H}]}|\mathcal{W}_\mathbb{H},W_{\tilde{M}+1:k-1},\mathbb{H},Q_{1:N}^{[k, \mathbb{H}]},	A_{1:n-1}^{[k, \mathbb{H}]}\right) \label{eq_L2_6}\\
&\qquad = \frac{1}{N} I \left(W_{k:K}; A_{1:N}^{[k, \mathbb{H}]}|\mathcal{W}_\mathbb{H},W_{\tilde{M}+1:k-1},\mathbb{H},Q_{1:N}^{[k, \mathbb{H}]} \right) \\
&\quad~\stackrel{\eqref{independency},\eqref{query_indep}}{=}
\frac{1}{N}I\left(W_{k:K};\mathbb{H},Q_{1:N}^{[k, \mathbb{H}]},A_{1:N}^{[k, \mathbb{H}]}|\mathcal{W}_\mathbb{H},W_{\tilde{M}+1:k-1} \right) \label{eq_L2_7}\\
&\qquad\stackrel{\eqref{reliability_constraint}}{=} \frac{1}{N} I \left(W_{k:K};W_k, \mathbb{H},Q_{1:N}^{[k, \mathbb{H}]}, A_{1:N}^{[k, \mathbb{H}]} |\mathcal{W}_\mathbb{H},W_{\tilde{M}+1:k-1} \right) - \frac{o(L)}{N} \label{eq_L2_8}\\
&\qquad = \frac{1}{N}  I \left(W_{k:K};W_k|\mathcal{W}_\mathbb{H},W_{\tilde{M}+1:k-1} \right) + \frac{1}{N}  I \left(W_{k:K};\mathbb{H},Q_{1:N}^{[k, \mathbb{H}]}, A_{1:N}^{[k, \mathbb{H}]}| \mathcal{W}_\mathbb{H},W_{\tilde{M}+1:k}\right)-\frac{o(L)}{N} \\
&\qquad = \frac{1}{N}I \left(W_{k+1:K};\mathbb{H},Q_{1:N}^{[k,\mathbb{H}]}, A_{1:N}^{[k, \mathbb{H}]}|\mathcal{W}_\mathbb{H},W_{\tilde{M}+1:k}  \right)+\frac{L-o(L)}{N}, \label{eq_L2_9}
\end{align}
where \eqref{eq_L2_1} follows from the non-negativity of mutual information, \eqref{eq_L2_2} follows from the privacy constraint, \eqref{eq_L2_3} follows from the independence of the messages and the queries, \eqref{eq_L2_4}, \eqref{eq_L2_6} follow from the fact that answer strings are deterministic functions of the messages and the queries, \eqref{eq_L2_5} follows from the fact that conditioning reduces entropy, \eqref{eq_L2_7} follows from the independence of $W_{k:K}$ and $\left(\mathbb{H}, Q_{1:N}^{[k, \mathbb{H}]}\right)$, \eqref{eq_L2_8} follows from the reliability constraint on $W_k$, and \eqref{eq_L2_9} follows from the independence of $W_k$ and $(\mathcal{W}_\mathbb{H},W_{\tilde{M}+1:k-1})$.
\end{Proof}

Now, we are ready to derive the lower bound for arbitrary $K$, $N$, and $\tilde{M}$. This can be obtained by applying Lemma~\ref{lemma_converse1} and Lemma~\ref{lemma_converse2} successively.
\begin{lemma}
For fixed $N$, $K$, and $\tilde{M} \leq M$, we have
\begin{align} \label{eq_L3}
D \geq L \left(1+ \frac{1}{N} + \dots + \frac{1}{N^{K-\tilde{M}-1}} \right) -o(L).
\end{align}
\end{lemma}
\begin{Proof}
We have
\begin{align}
D
& \stackrel{\eqref{eq_L1}}{\geq} L+ I\left(W_{\tilde{M}+2:K}; \mathbb{H}, Q_{1:N}^{[\tilde{M}+1, \mathbb{H}]}, A_{1:N}^{[\tilde{M}+1, \mathbb{H}]}|\mathcal{W}_{\mathbb{H}}, W_{\tilde{M}+1} \right) -o(L)    \label{c1}\\
& \stackrel{\eqref{eq_L2}}{\geq} L+ \frac{L }{N} +  \frac{1}{N}  I\left(W_{\tilde{M}+3:K};\mathbb{H}, Q_{1:N}^{[\tilde{M}+2, \mathbb{H}]}, A_{1:N}^{[\tilde{M}+2, \mathbb{H}]}|\mathcal{W}_{\mathbb{H}}, W_{\tilde{M}+1:\tilde{M}+2}  \right) -o(L)  \label{c2}\\
& \stackrel{\eqref{eq_L2}}{\geq} L+ \frac{L }{N} + \frac{L}{N^2}  +
\frac{1}{N}  I\left(W_{\tilde{M}+4:K};\mathbb{H}, Q_{1:N}^{[\tilde{M}+3, \mathbb{H}]}, A_{1:N}^{[\tilde{M}+3, \mathbb{H}]}|\mathcal{W}_{\mathbb{H}}, W_{\tilde{M}+1:\tilde{M}+3}  \right) -o(L) \\
& \stackrel{\eqref{eq_L2}}{\geq} \dots \\
& \stackrel{\eqref{eq_L2}}{\geq}  L \left(1+ \frac{1}{N} + \dots + \frac{1}{N^{K-\tilde{M}-1}} \right) - o(L), \label{c3}
\end{align}
where \eqref{c1} follows from Lemma~\ref{lemma_converse1}, \eqref{c2}-\eqref{c3} follow from applying Lemma~\ref{lemma_converse2} starting from $k=\tilde{M}+2$ to $k=K$, which differs from \cite{JafarPIR} in terms of the starting point of the induction.
\end{Proof}

We conclude the converse proof by dividing by $L$ and taking $L \rightarrow \infty$ in \eqref{eq_L3}, to have
\begin{align} \label{eq_lb}
D^* \geq 1+ \frac{1}{N} + \dots + \frac{1}{N^{K-\tilde{M}-1}}.
\end{align}
Finally, we note that the right hand side of \eqref{eq_lb} is monotonically decreasing in $\tilde{M}$. Since $\tilde{M}\leq M$, the lowest lower bound is obtained by taking $\tilde{M}=M$, which yields the final converse bound,
\begin{align} \label{eq_lb_f}
D^* \geq 1+ \frac{1}{N} + \dots + \frac{1}{N^{K-M-1}}.
\end{align}

\begin{remark}
We note that if \eqref{eq_lb_f} is tight, any prefetching strategy $\mathbf{m}$ such that $\sum_{n=1}^N m_n<M$ is strictly suboptimal. Furthermore, the lower bound in \eqref{eq_lb_f} is the same for all prefetching strategies $\mathbf{m}$ satisfying $\sum_{n=1}^N m_n=M$. In Section~\ref{achievability}, we show that this lower bound is tight.
\end{remark}

\section{Achievability Proof} \label{achievability}

We first note that the achievability scheme proposed in \cite{chen2017capacity} for the PIR problem with completely unknown private side information also works for the PIR problem with partially known private side information here. The PIR scheme in \cite{chen2017capacity} is based on MDS codes and consists of two stages. The first stage determines the systematic part of the MDS code according to the queries generated in \cite{JafarPIR}, which protects the privacy of the desired message, i.e., in the first stage, the user designs the queries such that no information is leaked about which message out of the $K$ messages is the desired one. In the second stage, the user reduces the number of the downloaded equations by downloading the parity part of the MDS code only. For the case of partially known private side information here, two privacy constraints should be satisfied: the desired message privacy constraint and the side information privacy constraint. For the desired message, we note that the user should guarantee that the queries designed to retrieve any of the $K-m_n$ messages should be indistinguishable at the $n$th database (i.e., with the exception of the $m_n$ messages that the $n$th database has provided). Due to the first stage, the privacy of the desired message holds as it was designed to protect the privacy of all $K$ messages, which is more restricted. Furthermore, the PIR scheme in \cite{chen2017capacity} also protects the privacy of the side information. The scheme in \cite{chen2017capacity} ensures that the queries do not reveal the identity of the $M$ messages that are possessed by the user as side information. In our model, we note that we need to protect the privacy of $M-m_n$ messages from the $n$th database, as the remaining $m_n$ messages are known to the $n$th database. Since the privacy constraint imposed on the side information in our model is less restricted than \cite{chen2017capacity}, using the scheme in \cite{chen2017capacity} satisfies the privacy constraint of the side information in our case as well. That is, the $n$th database cannot infer which other $M-m_n$ messages the user holds. The PIR scheme in \cite{chen2017capacity} achieves the normalized download cost in Theorem~\ref{thm1}. The PIR scheme in  \cite{chen2017capacity} requires a message size of $N^K$ symbols. In the following, we propose another achievability scheme which requires a message size of $N^{K-\frac{M}{N}}$, if  $m_n=\frac{M}{N} \in \mathbb{N}$. Thus, this scheme requires smaller sub-packetization and smaller field size for the MDS code.

Our PIR scheme for partially known private side information is based on the PIR schemes in \cite{JafarPIR, chen2017capacity}. To protect the privacy of the partially known private side information and the privacy of the desired message, similar to \cite{JafarPIR}, we apply the following three principles recursively: 1) database symmetry, 2) message symmetry within each database, and 3) exploiting undesired messages as side information. We reduce the download cost by utilizing the reconstruction property of MDS codes by exploiting partially known private side information as in \cite{chen2017capacity}. The side information enables the user to request reduced number of equations as a consequence of the user's knowledge of $M$ messages from the prefetching phase. Nevertheless, to protect the privacy of the side information, the user actually queries MDS coded symbols which is mixture of $K-m_n$ messages. The main difference between our achievability scheme and that in \cite{JafarPIR, chen2017capacity} is that since the $n$th database knows that the user has prefetched $m_n$ messages, the user does not need to protect the privacy for these $m_n$ messages from the $n$th database. This effectively reduces the number of messages that the scheme in \cite{chen2017capacity} needs to operate on to $K-m_n$ messages in contrast to $K$ in \cite{chen2017capacity}. When $\frac{M}{N}\in \mathbb{N}$, we show that if the user caches the same number of messages from each database, i.e., $m_n=\frac{M}{N}$, for all $n$, then the lower bound in \eqref{thm1_lb} is achievable by this scheme. This scheme reduces the message size requirement from $L=N^{K}$ in \cite{chen2017capacity} to $L=N^{K-\frac{M}{N}}$ here, simplifying the achievable scheme.

\subsection{Motivating Examples}
\subsubsection{$N=2$ Databases, $K=4$ Messages, and $M=2$ Cached Messages}
Assume that each message is of size $8$ symbols. We use $a_i$, $b_i$, $c_i$ and $d_i$, for $i=1,\ldots,8$,  to denote the symbols of messages $W_1$, $W_2$, $W_3$ and $W_4$, respectively. In this example, in the prefetching phase, the user caches message $W_3$ from database $1$, and message $W_4$ from database $2$; and in the retrieval phase, the user wishes to retrieve message $W_1$ privately. The user first generates the query table in Table~\ref{table_ex1}. In Table~\ref{table_ex1}, the user queries $7$ symbols. Since the user knows $d_1$ from the cached message $W_4$, in order to use the partially known private side information, the user can in fact reduce the number of queries to $6$ equations per database by ignoring $d_1$. However, if the user simply does not download $d_1$, it compromises the privacy of $W_4$ at database $1$. Alternatively, the user queries the MDS coded version of the $7$ symbols. By using these $7$ symbols as the systematic part, we can use a $(13,7)$ MDS code. By downloading the $6$ parity symbols, the user can reconstruct the whole $7$ symbols utilizing the knowledge of $d_1$. Therefore, the normalized download cost for our achievability scheme is $\frac{6+6}{8}=\frac{3}{2}$, which matches the lower bound in \eqref{thm1_lb} for this case.

For database $1$, the query table in Table~\ref{table_ex1} induces the same distribution on the messages $W_1$, $W_2$ and $W_4$. Therefore, we guarantee the privacy of the desired message. The reliability constraint can also be verified. Note that $b_2$ is downloaded from database $2$, and $d_2$ is downloaded in the prefetching phase. Therefore, $a_3$ and $a_4$ are decodable. By getting $b_4+c_3$ from database $2$, the user can get $b_4$ due to the private side information $W_3$. Therefore, the user can decode $a_7$ from $a_7+b_4+d_4$. Similar arguments follow for database $2$.

\begin{table}[h]
\caption{Query table for $K=4$, $N=2$, $M=2$.}
\centering
\begin{tabular}{cc}
	\hline
	\multicolumn{1}{|c|}{DB1}            & \multicolumn{1}{c|}{DB2}       			\\ \hline
	\multicolumn{1}{|c|}{$a_1$}          & \multicolumn{1}{c|}{$a_2$}     			\\ \hline
	\multicolumn{1}{|c|}{$b_1$}          & \multicolumn{1}{c|}{$b_2$}     			\\ \hline
	\multicolumn{1}{|c|}{$d_1$}          & \multicolumn{1}{c|}{$c_1$}     			\\ \hline
	\multicolumn{1}{|c|}{$a_3+b_2$}      & \multicolumn{1}{c|}{$a_5+b_1$}           \\ \hline
	\multicolumn{1}{|c|}{$a_4+d_2$}      & \multicolumn{1}{c|}{$a_6+c_2$}           \\ \hline
	\multicolumn{1}{|c|}{$b_3+d_3$}      & \multicolumn{1}{c|}{$b_4+c_3$}           \\ \hline
	\multicolumn{1}{|c|}{$a_7+b_4+d_4$}  & \multicolumn{1}{c|}{$a_8+b_3+c_4$}       \\ \hline
                                	  	 &                                          \\ \hline
	\multicolumn{1}{|c|}{$\mathcal{W}_{\mathbb{H}_1}$=\{$W_3$\}}  & \multicolumn{1}{c|}{$\mathcal{W}_{\mathbb{H}_2}$=\{$W_4$\}} \\\hline
\end{tabular}
\label{table_ex1}
\end{table}

\subsubsection{$N=2$ Databases, $K=5$ Messages, and $M=2$ Cached Messages}
Assume that each message is of size $16$ symbols. We use $a_i$, $b_i$, $c_i$, $d_i$ and $e_i$, for $i=1,\ldots,16$, to denote the symbols of messages $W_1$, $W_2$, $W_3$, $W_4$, and $W_5$, respectively. In this example, in the prefetching phase, the user caches message $W_4$ from database $1$, and message $W_5$ from database $2$; and in the retrieval phase, the user wishes to retrieve message $W_1$ privately. The user first generates the query table in Table~\ref{table_ex2}. In Table~\ref{table_ex2}, the user queries $15$ symbols. Since the user knows $e_1$ from the cached message $W_5$, in order to use the partially known private side information, the user in fact queries the MDS coded version of the $15$ symbols. By using these $15$ symbols as the systematic part, we can use a $(29,15)$ MDS code. By downloading the $14$ parity symbols, the user can reconstruct the whole $15$ symbols. Therefore, the normalized download cost for our achievability scheme is $\frac{14+14}{16}=\frac{7}{4}$, which matches the lower bound in \eqref{thm1_lb} for this case.

For database $1$, the query table in Table~\ref{table_ex2} induces the same distribution on the messages $W_1$, $W_2$, $W_3$ and $W_5$. Therefore, we guarantee the privacy of the desired message. The reliability constraint can also be verified. Note that $b_2$, $c_2$ are downloaded from database $2$, and $e_2$ is downloaded in the prefetching phase. Therefore, $a_3$, $a_4$ and $a_5$ are decodable. By getting $b_6+d_3$ from database $2$, the user can get $b_6$ due to the private side information $W_4$. Similarly, $c_6$ is also decodable. Therefore, the user can decode $a_{10}$ from $a_{10}+b_6+e_5$ and $a_{11}$ from $a_{11}+c_6+e_6$. By getting $b_8+c_8+d_7$ from database $2$, the user can get $b_8+c_8$ due to the private side information $W_4$. Therefore, the user can decode $a_{15}$ from $a_{15}+b_8+c_8+e_8$. Similar arguments follow for database $2$.

\begin{table}[h]
\caption{Query table for $K=5$, $N=2$, $M=2$.}
\centering
\begin{tabular}{cc}
\hline
\multicolumn{1}{|c|}{DB1}            		& \multicolumn{1}{c|}{DB2}				    \\ \hline
\multicolumn{1}{|c|}{$a_1$}          		& \multicolumn{1}{c|}{$a_2$}     			\\ \hline
\multicolumn{1}{|c|}{$b_1$}          		& \multicolumn{1}{c|}{$b_2$}     			\\ \hline
\multicolumn{1}{|c|}{$c_1$}          		& \multicolumn{1}{c|}{$c_2$}     			\\ \hline
\multicolumn{1}{|c|}{$e_1$}          		& \multicolumn{1}{c|}{$d_1$}     			\\ \hline
\multicolumn{1}{|c|}{$a_3+b_2$}      		& \multicolumn{1}{c|}{$a_6+b_1$}     		\\ \hline
\multicolumn{1}{|c|}{$a_4+c_2$}      		& \multicolumn{1}{c|}{$a_7+c_1$}     		\\ \hline
\multicolumn{1}{|c|}{$a_5+e_2$}      		& \multicolumn{1}{c|}{$a_8+d_2$}     		\\ \hline
\multicolumn{1}{|c|}{$b_3+c_3$}      		& \multicolumn{1}{c|}{$b_5+c_5$}     		\\ \hline
\multicolumn{1}{|c|}{$b_4+e_3$}      		& \multicolumn{1}{c|}{$b_6+d_3$}     		\\ \hline
\multicolumn{1}{|c|}{$c_4+e_4$}      		& \multicolumn{1}{c|}{$c_6+d_4$}     		\\ \hline
\multicolumn{1}{|c|}{$a_9+b_5+c_5$}     	& \multicolumn{1}{c|}{$a_{12}+b_3+c_3$}     \\ \hline
\multicolumn{1}{|c|}{$a_{10}+b_6+e_5$}  	& \multicolumn{1}{c|}{$a_{13}+b_4+d_5$}     \\ \hline
\multicolumn{1}{|c|}{$a_{11}+c_6+e_6$} 		& \multicolumn{1}{c|}{$a_{14}+c_4+d_6$}     \\ \hline
\multicolumn{1}{|c|}{$b_7+c_7+e_7$} 		& \multicolumn{1}{c|}{$b_8+c_8+d_7$}     	\\ \hline
\multicolumn{1}{|c|}{$a_{15}+b_8+c_8+e_8$} 	& \multicolumn{1}{c|}{$a_{16}+b_7+c_7+d_8$} \\ \hline
		                                    & 	                                     	\\ \hline
\multicolumn{1}{|c|}{$\mathcal{W}_{\mathbb{H}_1}$=\{$W_4$\}}& \multicolumn{1}{c|}{$\mathcal{W}_{\mathbb{H}_2}$=\{$W_5$\}} \\\hline
\end{tabular}
\label{table_ex2}
\end{table}

\subsection{General Achievable Scheme for $\frac{M}{N}\in \mathbb{N}$}

Let $\frac{M}{N}=m$. In the prefetching phase, the user caches $m$ messages from each database. To achieve the lower bound shown in \eqref{thm1_lb}, in the retrieval phase, we choose the message size as $L=N^{K-m}$ symbols. The details of the achievable scheme are as follows:
\begin{enumerate}
\item \textit{Initialization:} The user permutes each message randomly and independently. After the random permutation, we use $U_i(j)$ to denote the $j$th symbol of the permuted message $W_i$. Suppose the user wishes to retrieve $W_\theta$ privately. We then prepare the query table by first querying $U_\theta(1)$ from database $1$. Set the round index to $r=1$.

\item \textit{Symmetry across databases:} The user queries the same number of equations with the same structure as database $1$ from the remaining databases.   	

\item \textit{Message symmetry:} For each database, to satisfy the privacy constraint, the user should query equal amount of symbols from all other $K-m$ messages. Since the user has cached $m$ messages from each database in the prefetching phase, the user does not need to protect the privacy for these $m$ messages. For the $r$th round, the user queries sums of every $r$ combinations of the $K-m$ messages.

\item \textit{Exploiting side information:} For database $1$, the user exploits the side information equations obtained from the other $(N-1)$ databases to query sum of $r+1$ combinations of the $K-m$ messages, where sum of $r$ combinations is the side information. If the $r$ combinations contain the cached message from database $1$, we replace the overlapping symbols through the symbols cached from other databases.

\item \textit{Repeat} steps 2, 3, 4 after setting $r=r+1$ until $r=K-m+1$.

\item \textit{Shuffling the order of queries:} By shuffling the order of queries uniformly, all possible queries can be made equally likely regardless of the message index. This guarantees the privacy of the desired message.

\item \textit{Downloading MDS parity parts:} Now, the query table is finished. For each database, let $p$ be the number of queried symbols in the query table, and let $q$ be the number of queried symbols which are determined by the side information the user cached in the prefetching phase. Apply a $(2p-q, p)$ MDS code to the queried symbols by letting the $p$ symbols to be the systematic part. Finally, the user downloads the parity parts of the MDS-coded answering strings which are $p-q$ symbols for each database.
\end{enumerate}	

\subsection{Normalized Download Cost}
We now calculate the total number of downloaded symbols. We first calculate $p$, which is the number of queried symbols in the query table for each database,
\begin{align}
p&=\binom{K-m}{1}+\binom{K-m}{2}(N-1)+\dots + \binom{K-m}{K-m}(N-1)^{K-m-1}  \label{eq_p} \\
 &=\frac{1}{N-1} \left[ \binom{K-m}{1}(N-1) +\binom{K-m}{2}(N-1)^2 + \dots +\binom{K-m}{K-m}(N-1)^{K-m}  \right] \\
 &=\frac{1}{N-1} \left( N^{K-m} -1  \right),
\end{align}
where $\binom{K-m}{r}$ in \eqref{eq_p} corresponds to the queries of sums of every $r$ combinations of the $K-m$ messages, and $(N-1)^{r-1}$ corresponds to the number of sets of the available side information from other $(N-1)$ databases.

We then calculate $q$, which is the number of queried symbols which are determined by the side information the user cached in the prefetching phase,
\begin{align}
q&=\binom{(N-1)m}{1} + \binom{(N-1)m}{2} (N-1) + \dots + \binom{(N-1)m}{(N-1)m} (N-1)^{(N-1)m-1}  \label{eq_q} \\
 &=\frac{1}{N-1} \left[ \binom{(N-1)m}{1} (N-1) +\dots + \binom{(N-1)m}{(N-1)m}(N-1)^{(N-1)m} \right] \\
 &=\frac{1}{N-1} \left( N^{(N-1)m} -1             \right),
\end{align}
where $\binom{(N-1)m}{r}$ in \eqref{eq_q} corresponds to the queries which can be determined by the partially known private side information, and $(N-1)^{r-1}$ corresponds to the number of sets of queries consisting of $r$ combinations.

Next, we calculate the number of symbols for the desired message,
\begin{align}
L&=N \left[\binom{K-m-1}{0} + \binom{K-m-1}{1}(N-1) +\dots + \binom{K-m-1}{K-m-1}(N-1)^{K-m-1}\right] \label{eq_L} \\
&=N \times N^{K-m-1} = N^{K-m},
\end{align}
where $\binom{K-m-1}{r-1}$ in \eqref{eq_L} corresponds to the queries containing the desired message and $(N-1)^{r-1}$ corresponds to the number of sets of queries consisting of $r$ combinations.

Therefore, the normalized download cost becomes,
\begin{align}
\frac{D}{L}&= \frac{N(p-q)}{L} \\
&=\frac{\frac{N}{N-1} \left( N^{K-m} -1  \right)-  \frac{N}{N-1} \left( N^{(N-1)m} -1             \right)  }{N^{K-m}} \\
&=\frac{N}{N-1} \times \frac{N^{K-m}-N^{(N-1)m} }{ N^{K-m}} \\
&=\frac{1}{1-\frac{1}{N}} \times \left[1- \left(\frac{1}{N}\right)^{K-M} \right],
\end{align}
which matches the lower bound in \eqref{thm1_lb}.

\begin{remark}
Note that although our achievable scheme and the scheme in \cite{chen2017capacity} are both using MDS coding to exploit the available side information, the field size requirements for realizing the MDS codes are different. For the scheme of \cite{chen2017capacity}, a $(2\tilde{p}-\tilde{q},\tilde{p})$ MDS code is used, where $\tilde{p}=\frac{1}{N-1}(N^{K}-1)$ and $\tilde{q}=\frac{1}{N-1}(N^{M}-1)$. This requires larger field size than the $(2p-q,p)$ MDS code used in our scheme (if $\frac{M}{N} \in \mathbb{N}$), since $2\tilde{p}-\tilde{q}> (2p-q)$.
\end{remark}

\section{Conclusion}
In this paper, we have introduced a new PIR model, namely, PIR with partially known private side information as a natural model for studying practical PIR problems with cached side information. In this model, the user and the databases engage in a caching/PIR scenario which consists of two phases, namely, prefetching phase and retrieval phase. The $n$th database provides the user with $m_n$ side information messages in the prefetching phase such that $\sum_{n=1}^N m_n \leq M$, hence, each database has {\it partial knowledge} about the side information in contrast to full knowledge in \cite{tandon2017capacity} and no knowledge in \cite{wei2017fundamental, chen2017capacity, kadhe2017private}. Based on this side information, the user designs a retrieval scheme that does not reveal the identity of the desired message or the identities of the remaining $M-m_n$ messages to the $n$th database. For this model, we determined the exact capacity to be $C=\frac{1-\frac{1}{N}}{1-(\frac{1}{N})^{K-M}}$. The capacity is attained for any prefetching strategy that satisfies the cache memory size constraint with equality. The achievable scheme in \cite{chen2017capacity} can also be used for this model. We further proposed another PIR scheme which requires smaller sub-packetization and field size for the case of uniform prefetching. Uniform prefetching, when feasible, is optimal. Interestingly, the capacity expression we derive for this problem is exactly the same as the capacity expression for the PIR problem with completely unknown side information \cite{chen2017capacity}. Therefore, our result implies that there is no loss in employing the same databases for prefetching and retrieval purposes.

\bibliographystyle{unsrt}
\bibliography{references}

\end{document}